\documentstyle[12pt,epsf,psfig,epsfig]{laa}     

\newcommand{\dgr}{$^{\circ}$}
\newcommand{\cep}{Cepheus X-4 }

\newcommand{\Halp}{H$_{\alpha} $ }
\newcommand{\Hbet}{H$_{\beta} $ }
\newcommand{\Hgam}{H$_{\gamma}{ }$ }

\begin{document}                                          
\thesaurus{02.12.2, 08.02.3, 08.05.2, 08.16.7 Cep X-4, 09.04.1, 13.25.5}
\title{The identification of the transient X-ray pulsar Cepheus X-4
with a Be/X-ray binary.
\thanks{Based on observations collected at the Observatoire de Haute-Provence,
CNRS, France}}
\author{J.M. Bonnet-Bidaud\inst{1}
\and M. Mouchet\inst{2,3}} 
\offprints{J.M. Bonnet-Bidaud}
\institute
{Service d'Astrophysique, CEA, DSM/DAPNIA/SAp, CE-Saclay, 
F-91191 Gif sur Yvette Cedex, France \\
 email:bobi@sapvxg.saclay.cea.fr
\and
DAEC, Observatoire de Paris, Section de Meudon, F-92195 Meudon Cedex, France,
email:mouchet@obspm.fr
\and
Universit\'e Denis Diderot, 2 Place Jussieu, F-75005 Paris, France}
\date{Received date: November 28th, 1997; accepted date: January 20th, 1998}
\maketitle

\begin{abstract}
We present long slit spectral (375-725 nm) 
observations  of the proposed identification of the transient 
66s X-ray pulsar Cep X-4=GS2138+56. Spectra show features typical of Be/X-ray binaries.
Superimposed on a weak emission from the IC 1396 nebula, strong H$\alpha$ 
(4.5 nm EW) and H$\beta$ (0.3 nm EW) lines are seen in emission with the 
other Balmer lines in absorption. 
Significant interstellar absorption features are also detected, including a 
strong Na I doublet (589 nm) and diffuse bands at 443, 578 and 628 nm.
From the shape of the continuum as well as the lines present in the 
spectrum, a most probable spectral type of B1-B2V is derived with a reddening
of  E$_{B-V}$=1.3$\pm$0.1. The reddening value is corroborated by
the measure of interstellar absorption features, except for 
the sodium line which appears to be in excess and may be partly 
from circumstellar origin.
The optical absorption is fully consistent with the column density 
derived from X-ray spectra, therefore confirming the identification.
Despite apparent spatial coincidence, the source is located much further 
away than the local intervening nebula IC1396. 
The best estimate of the distance is D=(3.8$\pm$0.6) kpc, which places the
source in the outer Perseus arm of the Galaxy. 
At this distance, 
the X-ray quiescent luminosity is (3-6) 10$^{33}$ erg.s$^{-1}$, 
thus comparable to typical Be/X-ray binary low states.

\keywords{X-rays: binaries, Stars: emission line Be,  
Stars: pulsars: individual: Cep X-4}
\end{abstract} 

\section{Introduction}
The X-ray source Cep X-4 was discovered in a transient high level 
X-ray flux in 1972 by the satellite OSO-7 (Ulmer et al. 1973).
Early positions, accurate only to a fraction of degree, led to a 
suggested identification with the massive O6.5V star HD 206267, 
responsible for the excitation of the IC 1396 nebula (Liller 1974),
though further observations did not confirm 
the identification (Hensberge \& Hammerschlag 1975).
The source was not detected again by an X-ray satellite till the  
observation of a new outburst by GINGA in April 1988.
From these observations, coherent 66s pulsations were reported, 
revealing an X-ray pulsar 
with a complex X-ray spectrum, including a possible 30keV cyclotron 
absorption feature (Koyama et al. 1991, Mihara et al. 1991).
The position of the pulsar Cep X-4/GS2138+56 was refined to a 
(15'x6') error box and, assuming circularity,
Doppler changes of the X-ray pulses were used to constrain the 
orbit to a period more than 23 days, suggesting a massive
X-ray binary. 
A new outburst, detected in 1993 by the ROSAT satellite 
during pointed observations and a residual faint emission when the 
source has declined by a factor more than 1000, were used to further 
refine the position (Schulz, Kahabka \& Zinnecker 1995).
An association with a m$_{V}$=14.2 Be star at $\alpha_{2000}$ = 21h39m30.6s and
$\delta_{2000}$ = +56\dgr59'12.9" was proposed by Roche, Green \& Hoenig (1997)
on the basis of the strong H$\alpha$ emission (see also Argyle 1997).
We presented here the first detailed spectroscopic information on this star,
establishing the identification with a typical Be/X-ray  binary.
Preliminary results were given in Bonnet-Bidaud \& Mouchet (1997).

\section{Observations and results}
Spectroscopic observations were conducted at  
the 1.93m telescope  of the Haute-Provence Observatory (France),
equipped with the Carelec spectrograph (Lemaitre et al. 1990),
during two subsequent nights on July 29th and 30th, 1997.
Series of (375-725 nm) low resolution spectra were recorded 
through a long slit (2.3" x 5') 
aperture with typical exposure times of 15 min, covering 1 hr and 2.5 hr, 
respectively during the two nights. 
Calibration spectra using an helium lamp were recorded at 1 hr interval and
before and after the observations, from which a typical resolution of
1.3 nm was measured. 
 The standard spectroscopic stars BD+33 2642 and 
BD+28 4211 were observed during the two nights. The transparency was affected
by residual absorption during part of the nights and the resulting variability
was corrected by monitoring two additional stars present in the slit.

\begin{figure*}
\epsfig{file=gsfig1.ps,clip=,width=18.0cm,bbllx=39,bblly=333,bburx=556,bbury=508}
\caption[ ]{Optical identification of the transient X-ray pulsar \cep.\\ 
a- The (2x2)\dgr{ }optical field around the source showing the intervening
nebula IC1396. 
b- The (30x30 arcmin) ROSAT (0.1-2.5keV) X-ray image during quiescence. 
{ }c- The (10x10 arcmin) optical image around the X-ray source from 
the digitized Palomar Sky Survey.} 
\end{figure*}

\begin{table*}
\caption{Characteristics of the H$\alpha$ bright Be/X-ray binaries} 
\smallskip
\small
\begin{tabular}{rrrrrrrrll}
\multicolumn{1}{c}{\rm Source } &
\multicolumn{3}{c}{\rm Emission line EW (nm)}  &
\multicolumn{2}{c}{\rm Interstellar line EW (nm)}   & 
\multicolumn{1}{c}{\rm E$_{B-V}$} & 
\multicolumn{1}{c}{\rm D}    &  
\multicolumn{1}{c}{\rm Spectral type } & \multicolumn{1}{l}{\rm References} \\
\multicolumn{1}{c}{\em  } &
\multicolumn{1}{c}{\em {H$\alpha^{max}$ }} & 
\multicolumn{1}{c}{\em {H$\alpha$}} & \multicolumn{1}{c}{\em H$\beta$} &
\multicolumn{1}{c}{\em NaD}         & \multicolumn{1}{c}{\em DIB 443 nm} &
\multicolumn{1}{c}{\em  } &
\multicolumn{1}{c}{\em kpc} & 
\multicolumn{1}{c}{\em  }  & \multicolumn{1}{c}{\em  } \\
\hline
\multicolumn{10}{l}{} \\
 GS2138+56 & 5.3 & 4.54 (11) & 0.33 (2)  & 0.32 (3) & 0.28 (2) & 1.1-1.2 & 3.8(6) & B1V-B2V       & \\
 A1118-61  & 8.9 & 5.4-6.7   & 0.45-0.67 & 0.24 (3) & 0.27 (5) & 1.2(2)  & 5 (2)  & O9.5III-V & 1,2,3\\
 4U1145-61 & 4.5 & 1.1-1.5   & 0.15-0.45 &  -       &  -       & 0.29    & 3.1(5) & B1V     & 1,4\\
 A0535+25  & 3.0 & 1.0-1.9   & 0.1-0.2   & 0.11(2)  & 0.27 (5) & 0.75    & 2.6(4) & O9.7V    & 1,5,6\\
\hline
\multicolumn{10}{l}{Error bars on last digits are given into parentheses } \\
\multicolumn{10}{l}{$^{1}$Reig et al.1997, $^{2}$Motch et al.1988, $^{3}$Mouchet (unpublished),
$^{4}$Stevens et al.1997, $^{5}$Janot-Pacheco et al.1987, $^{6}$Wade \& Oke 1977} \\
\end{tabular}
\end{table*}

Figure 1 shows a (2x2)\dgr{ }field-of-view around the position of the 
X-ray source Cep X-4. The field is rather complex, due to the presence of
the low surface brightness nebula, IC1396, excited by the O6 Trapezium-type
system, HD206267 (star 11), itself included in a very young open
cluster, Trumpler 37, at a distance modulus of 10.0 (Marschall et al. 1990). 
The X-ray source is located in the southern part of the
nebula, 4.1 arcmin  north-east from the bright star HD239725 (star 12).
The optical image corresponding to the ROSAT field-of-view is shown in 
Figure 1c with the identification first proposed by Roche et al. (1997) 
at the center of the field. Star 12, also detected by ROSAT, is the nearest 
X-ray source (see fig 1b).\\
The mean spectrum of the suggested counterpart of \cep is shown
in Figure 2. It shows an heavily reddened continuum with a strong \Halp
line and a weaker \Hbet line. No other strong emission lines are seen, the other 
most conspicuous features are atmospheric and interstellar absorption lines.
The sky spectrum, also shown in Fig. 2, reveals strong emission lines, including 
\Halp and \Hbet. It is representative of the rather uniform emission from 
the excited IC1396 nebula, seen through the entire 5' aperture.
The very accurate sky-line subtraction demonstrates that the extracted spectrum 
is not contaminated by the foreground nebula.\\
The normalized spectrum is shown in Fig. 3. 
Except for the \Halp and \Hbet lines in emission, the rest of the
Balmer series is seen in absorption down to H$_{9}$. The \Hgam line, 
not clearly detected, is the transition from emission to absorption. 
Such characteristics are usually found in Be stars.
Strong absorption features are also visible, including a deep narrow doublet 
NaD at 589.3 nm (not resolved at the present resolution) and several diffuse 
interstellar bands (DIB) at 443.0 nm, 577.7 nm, 618.0 nm and 628.3 nm.
Other bands are telluric O$_{2}$ band at 630.0 and 688.4 nm and atmospheric vapour
band at 718.5 nm. We also identify the Fe multiplet complex in
emission at 501.7, 517.0 and 531.5 nm, as well as shortward of \Hbet at
462.8 and 472.8 nm. The only helium line clearly present in this spectrum is
HeI at 447.1 nm on the edge of the 443.0 nm DIB and possibly at 471.3 nm.
The HeII 468.6 nm, typical of the X-ray heated stars, is not detected
nor any significant excess around the CIII-NIII blend (465nm). 
Several features are not clearly identified, such as a bump
around 560 nm and an absorption near 647 nm. Absorption redward of \Hbet
could be an inverse P-Cygni profile sometimes observed in Be stars or
a CrI line as suggested for another Be/X-ray binary
A1118-61 by Polcaro et al. (1993), but also more likely the 488.5 nm DIB.\\
Equivalent widths (EW) of the most conspicuous lines are reported in Table 1.
Spectral variability has been searched both in the continuum and lines.
The \Halp and \Hbet EW were found only slightly variable by 4.4\% and 2.5\%
respectively, while the continuum varies by less than 10 \% over the two days
of observations.
The \Halp is broad with a deconvolved FWHM of (530$\pm$40) km.s$^{-1}$. 
Using the empirical relation between the width and EW of the \Halp line
derived by Dachs et al. (1986), this 
corresponds to a projected rotational velocity of (460$\pm$50) km.s$^{-1}$.
\\

\begin{figure}
\epsfig{file=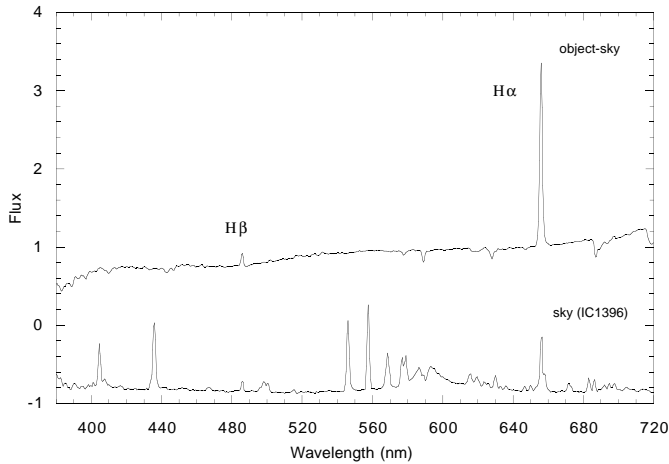,clip=,width=8.8cm,bbllx=48,bblly=265,bburx=552,bbury=615}
\caption[ ]{The mean spectrum of the \cep optical counterpart.
Flux after airmass correction, sky subtraction and 
calibration is shown but without correction for the interstellar reddening. 
Flux units are 10$^{-22}$ W.m$^{-2}$.nm$^{-1}$.
Also shown is the mean sky spectrum revealing the presence of the 
ionized HII region through which the source is seen.  The sky spectrum, averaged over 0.1 arcmin$^{2}$,
have been shifted downward by one unit for clarity.}
\end{figure}

\begin{figure}
\epsfig{file=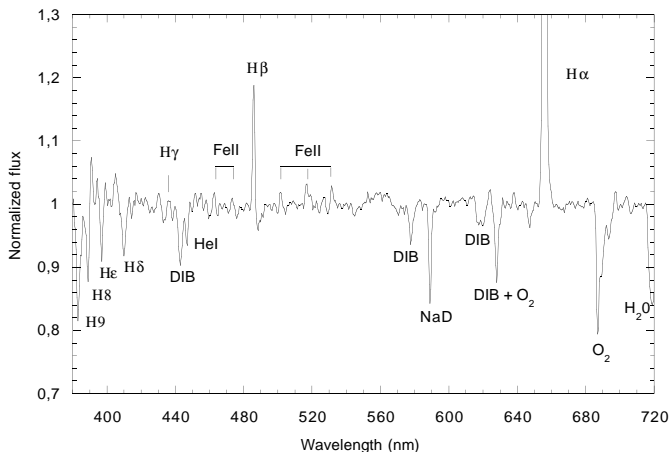,clip=,width=8.8cm,bbllx=34,bblly=265,bburx=552,bbury=614}
\caption[ ]{The source normalized mean spectrum.}
\end{figure}

\section{Discussion}
The spectrum of the suggested \cep optical counterpart appears typical of a Be star
with a strong \Halp line in emission and a Balmer series turning to absorption 
shortward of \Hgam. The probability to find such an active star inside the small
(15'x6') error box of an X-ray source is sufficiently low to consider that the object is
indeed the optical counterpart of the 66s X-ray pulsar \cep.
The significance of the coincidence may be altered as the
source is located in the line of sight of the open cluster, Trumpler 37,
at the center of an OB association, but a Be star located in the cluster
will appear much more luminous (see below). The high excitation 
HeII (468.6 nm) line is not always present in X-ray stars and 
its absence here may results from an X-ray low state.\\
Based on the compilation by Reig et al. (1997), we note that the star is 
the second brightest in \Halp among other identified Be/X-ray binaries (see Table 1).
Assuming the approximate linear relation suggested by Reig et al. (1997) between \Halp
EW and the orbital period, yields an orbit of the order of 130-170 days, of the same 
order than the $\sim$ 100 days period derived from the P$_{spin}$-P$_{orb}$ correlation 
(Koyama et al. 1991).

\subsection{Spectral type of the \cep optical counterpart} 
Significant constraints on the system can be derived from the shape of the
continuum. A sample of representative spectra from O8 to B7 stars among 
classes III and V was selected from the library of Jacoby et al. (1984) and 
the continuum in regions free of lines was fitted,
using standard $\chi^{2}$ method, to the observed source spectrum after correcting for 
interstellar absorption. Dereddening was applied according to the mean galactic law 
of Seaton (1979) with values of E$_{B-V}$ ranging from 0 to 2.0.\\
Figure 4 shows the result of the fit. The best agreement is obtained for 
 E$_{B-V}$=1.3$\pm$0.1, where the range corresponds to 1$\sigma$ error
bars for a 2dim-$\chi^{2}$ fit. The fit only loosely
constrains  the spectral type between 09V and B4V.
However, additional clues on the nature of the star are also given by the line spectrum.
The absence of the HeII Pickering series is consistent with a spectral type at or
later than B0 (Walborn and Fitzpatrick 1990) while the presence of the HeI lines
in absorption constrains to types earlier than B5-B7 (Collins \& Sonneborn 1977).
The HeI lines may however be affected by the veiling effect, a filling of 
the underlying absorption spectrum by the emission from the envelope.
On the other hand, the FeII lines are clearly seen, which, according to Jaschek et al.
(1980) is expected from spectral types peaking at B2. This is also in accordance with 
the absence of significant SiII (412.8 nm) and MgII (448.1 nm) lines whose 
contribution increases towards types later than B2 (Walborn \& Fitzpatrick 1990),
though the moderate resolution somewhat limits this conclusion. The most probable 
spectral type is therefore a B1-B2V star.

\begin{figure}
\epsfig{file=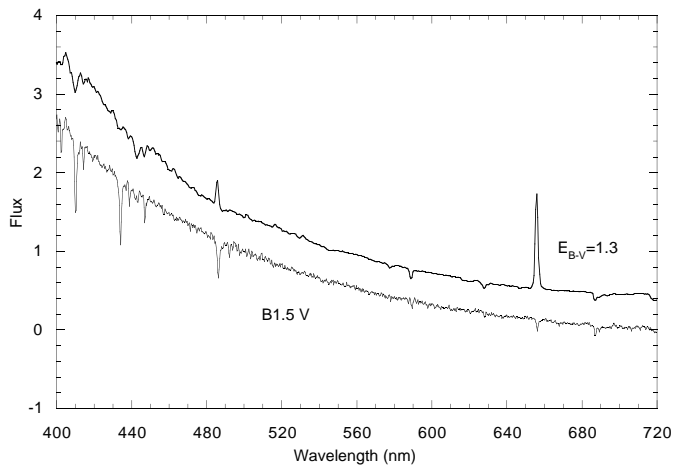,clip=,width=8.8cm,bbllx=54,bblly=523,bburx=442,bbury=794}
\caption[ ]{Best fit of the observed optical continuum. The source has been corrected 
for absorption with E$_{B-V}$=1.3 and a B1.5V representative spectrum is shown shifted 
by 0.4 unit for clarity.}
\end{figure}

\subsection{Absorption towards \cep}        
A significant amount of interstellar absorption is visible in the spectrum with 
a strong NaD line with EW=0.32$\pm$0.03 nm and pronounced diffuse interstellar bands, 
including the DIB (443 nm) with EW = 0.28$\pm$0.02 nm (see Table 1). Based on observations
of stars with E$_{B-V}$ ranging from 0 to 1.5, a nearly linear relation
between the EW of the DIB(443 nm) and the colour excess E$_{B-V}$ has been derived 
by various authors. From the observed DIB EW, values of E$_{B-V}$=1.1-1.3 and
E$_{B-V}$=1.0-1.2 are derived, using relations by Herbig (1975) and T\"ug \&
Schmidt-Kaler (1981) respectively. These values are consistent with the one derived from
the continuum fit. \\
A discrepant result is however obtained if one uses the NaD line as a tracer of
interstellar absorption. Using the parabolic relation 
NaD1/E$_{B-V}$, recently derived by Munari \& Zwitter (1997) from observations
of nearby stars, and assuming equal contribution from each component of the doublet, 
a value of E$_{B-V} \sim$ 1.3 would correspond to a NaD EW of $\sim$ 0.15 nm, 
approximatively half the observed value. The observed NaD in \cep is indeed the highest
among the sources in Table 1, irrespective of the distance. This could result from
peculiar abundances in the interstellar medium but also more likely from possible
circumstellar contribution. Circumstellar sodium has been reported in the Be 
shell star
$\zeta$ Tau by Delplace (1970) with EW up to 0.26 nm. The observed excess 
of EW$\sim$0.17 nm, could possibly have the same origin. 
 We note that such absorption is usually detected mostly among 
Be "shell" stars. The high rotational velocity derived from the \Halp line width 
(460 km.s$^{-1}$) is close to the break-up velocity (470-520) km.s$^{-1}$ 
of B1-B2V stars (Collins \& Sonneborn 1977), so that
an inclination of the equatorial plane greater than (65-80)\dgr{ } is indicated, which
could explain the similarity of  \cep  with the Be-shell class.\\

\subsection{Location in the Galaxy}
Reddening  towards the star HD239725 
(star 12), in the foreground open cluster Trumpler 37, 
at a projected small distance of the source, has been measured 
to E$_{B-V}$=0.46 (Clayton \& Fitzpatrick 1987), so that the \cep
counterpart is clearly located behind the cluster. 
Assuming a standard gas-to-dust ratio (Ryter et al. 1975), 
the reddening implies
a N$_{H}$ column density in the direction of the source of
N$_{H}$ = (8.8$\pm$0.8) 10$^{21}$ at.cm$^{-2}$. This value is in  
good agreement with the X-ray column density of 
N$_{H}$ = (7.9$\pm$1.9) 10$^{21}$ at.cm$^{-2}$ derived by 
Schulz et al. (1995) from X-ray spectra. 
With a mean interstellar density n${_e}<$ 1 at.cm$^{-3}$, this
corresponds to a distance greater than (2.9$\pm$0.2) kpc.
At a galactic position of (l$^{II}$=99.01\dgr, b$^{II}$=+3.31\dgr), the source is 
located in the plane of the Galaxy in the direction of the outer Perseus
arm. The total column density of neutral hydrogen at this position is
N$_{HI}$ $\sim$ 9.4 10$^{21}$ at.cm$^{-2}$  (Dickey \& Lockman 1990) so
that the source is close to the most outer part of the Galaxy. 
In this direction, the Leiden-Greenbank HI survey (Burton 1985) clearly
reveals distinct galactic structures at v$\sim$-10 km.s$^{-1}$  and 
v$\sim$-80 km.s$^{-1}$. Based on a detailed HI study of this region, Simonson  
\& van Someren Greve (1976) concluded to the superposition of three
separate galactic contributions, a local star population at $\sim$ 400 pc,
the Cep OB2 association including the IC1396 nebula at $\sim$ 830 pc and
a group of luminous stars at 3-5 kpc which corresponds to the Perseus
arm. In view of this galactic structure and of the distance derived from
reddening, it is most likely that the \cep counterpart belongs to this
last group.\\ 
The observed m$_{V}$=14.2 magnitude with a reddening of  E$_{B-V}$$\sim$1.3
and an absolute magnitude of Mv=-(2.5-3.8), corresponding to B1V-B2V stars 
(Balona \& Crampton 1974) as indicated by our spectral results, 
 corresponds to a distance D=(3.8$\pm$0.6) kpc in accordance with such location.

With the lack of distance determination and the spatial coincidence with the
Cep OB2 association, Schulz et al. (1995) have suspected that the
X-ray source may be located at D$<1$kpc, leading to the weakest observed
quiescence luminosity among Be/X-ray binaries. Our distance determination
implies instead a luminosity in quiescence, 
Lx(0.1-100keV)=(3.3-6.2) 10$^{33}$ erg.s$^{-1}$ and in outburst 
Lx=(0.6-1.2) 10$^{37}$ erg.s$^{-1}$, in the normal range of what is observed 
among typical Be/X-ray binaries (Apparao 1994).

In conclusion, optical spectra of the \cep transient X-ray pulsar appear 
fully consistent with a Be/X-ray system located at a distance of
D$\sim$4kpc, in the outer Perseus arm of the Galaxy. The optical spectrum 
indicates a B1-B2V companion showing one of the strongest \Halp
emissions among Be/X-ray binaries and a significative excess
of absorption in the sodium line with respect to what expected from
a purely interstellar origin. This and the determination of the radial 
velocity curve clearly deserve further optical studies. 

\medskip
\small
\it Acknowledgements. \rm We wish to thank P. Roche for early discussions 
on the source and M. Floquet, A.M. Hubert and E. Janot-Pacheco for valuable
informations on Be stars.

\end{document}